\documentclass[aps,twocolumn,showpacs]{revtex4}
\usepackage{graphicx,amsmath}
\newcommand{\figwidth}{1.0\columnwidth}
\newcommand{\eq}[1]{Eq.(\ref{#1})}
\newcommand{\fig}[1]{Fig.~\ref{#1}}
\newcommand{\tab}[1]{Table~\ref{#1}}
\newcommand{\sect}[1]{Section~\ref{#1}}
\newcommand{\avg}[1]{ {\langle #1 \rangle} }

\newcommand{\ahum}[1]{``#1''}

\newcommand{\ple}{P_{L,\epsilon}(\lambda)}
\newcommand{\plen}{P_{L,\epsilon}(E)}

\begin{document}

\title{Finite-size effects at first-order isotropic-to-nematic 
transitions}

\author{J.M. Fish and R. L. C. Vink}

\affiliation{Institute of Theoretical Physics, 
Georg-August-Universit\"at G\"ottingen, Friedrich-Hund-Platz~1, 37077 
G\"ottingen, Germany}

\date{\today}

\begin{abstract} We present simulation data of first-order isotropic-to-nematic 
transitions in lattice models of liquid crystals and locate the thermodynamic 
limit inverse transition temperature $\epsilon_\infty$ via finite-size scaling. 
We observe that the inverse temperature of the specific heat maximum can be 
consistently extrapolated to $\epsilon_\infty$ assuming the usual $\alpha / L^d$ 
dependence, with $L$ the system size, $d$ the lattice dimension and 
proportionality constant $\alpha$. We also investigate the quantity 
$\epsilon_{L,k}$, the finite-size inverse temperature where $k$ is the ratio 
of weights of the isotropic to nematic phase. For an optimal value $k 
= k_{\rm opt}$, $\epsilon_{L,k}$ versus $L$ converges to $\epsilon_\infty$ 
much faster than $\alpha/L^d$, providing an economic alternative to locate the 
transition. Moreover, we find that $\alpha \sim \ln k_{\rm opt} / {\cal 
L}_\infty$, with ${\cal L}_\infty$ the latent heat density. This suggests that 
liquid crystals at first-order IN transitions scale approximately as $q$-state 
Potts models with $q \sim k_{\rm opt}$.\end{abstract}



\maketitle

\section{Introduction}

The investigation of the isotropic-to-nematic (IN) transition in liquid 
crystals via computer simulation is long established. Decades 
ago Lebwohl and Lasher (LL) introduced a simple lattice model, the LL 
model, to study this transition \cite{physreva.6.426}. At each 
site $i$ of a cubic lattice they attached a three-dimensional unit 
vector $\vec{d}_i$ (spin) interacting with its nearest-neighbors via
\begin{equation}\label{eqll}
 {\cal H} = -\epsilon \sum_{<ij>} | \vec{d}_i \cdot \vec{d}_j |^p,
\end{equation}
where $p=2$ and with a factor $1/k_B T$ absorbed into the coupling constant 
$\epsilon>0$, with $k_B$ the Boltzmann constant and $T$ the temperature. Despite 
its simplicity, the LL model captures certain aspects of liquid crystal phase 
behaviour remarkably well, and has consequently received considerable attention 
\cite{citeulike:4197190, citeulike:4197214}.

A common problem in locating the IN transition via simulation is the 
issue of finite system size. Phase transitions are defined in the 
thermodynamic limit, whereas simulations always deal with finite 
particle numbers. In order to estimate the thermodynamic limit 
transition point, it is typical to perform a number of simulations for 
different system sizes and to subsequently extrapolate the results 
following some finite-size scaling (FSS) procedure. Which procedure to 
use depends on the type of transition, i.e.~whether it is continuous or 
first-order. In three spatial dimensions, the IN transition is typically 
first-order; in two dimensions both continuous \cite{bates.frenkel:2000, 
physreva.31.1776, citeulike:3740617} and first-order \cite{vink:2006*b, 
vink.wensink:2007} IN transitions can occur, depending on the details of 
the interactions \cite{physrevlett.52.1535, physrevlett.89.285702, 
enter.romano.ea:2006}. In this paper we focus on the first-order case.

The literature on FSS at first-order transitions is quite extensive, for a 
review see \cite{citeulike:1920630}, the majority of which deals exclusively 
with the Potts model \cite{note1}. An important result is that 
the \ahum{apparent} transition inverse temperature $\epsilon_{L,CV}$, obtained 
in a finite system of size $L$, is shifted from the thermodynamic limit value 
$\epsilon_\infty$ as \cite{citeulike:3716330, citeulike:3720308}
\begin{equation}\label{eq:lcv}
 \epsilon_{L,CV} = \epsilon_\infty - \alpha / L^d 
 + {\cal O}( 1/L^{2d} ),
\end{equation}
with proportionality constant $\alpha>0$. Here $\epsilon_{L,CV}$ is the 
inverse temperature where the specific heat in a finite system of size 
$L$ attains its maximum, $d$ is the spatial dimension of the lattice 
and $L$ denotes the linear extension of the simulation box, generally 
square or cubic, with periodic boundary conditions.

We emphasize that \eq{eq:lcv} was derived for the Potts model where the 
proportionality constant is known to be
\begin{equation}\label{eq:alpha}
 \alpha_{\rm Potts} = \ln q / {\cal L}_\infty,
\end{equation}
with $q$ the number of Potts states and ${\cal L}_\infty$ the 
latent heat density in the thermodynamic limit \cite{citeulike:3716330, 
citeulike:3720308}. Interestingly, simulations of the LL model have shown that 
the functional form of \eq{eq:lcv} also works well for IN transitions
\cite{physrevlett.69.2803, citeulike:3740162}. That is, meaningful 
extrapolations of $\epsilon_{L,CV}$ can be performed, although the significance 
of $\alpha$ is not obvious. It certainly cannot be related to the number of spin 
states, i.e.~conform to \eq{eq:alpha}, since the LL model is a {\it continuous} 
spin model, in contrast to the {\it discrete} spin variables of the Potts model 
\cite{revmodphys.54.235}.

In any case, based on the success of \eq{eq:lcv} in describing finite-size 
effects in the LL model, it could be hoped that other scaling relations, originally 
derived for the Potts model, also remain valid. Of particular interest is the 
result of Borgs and Kotecky, who showed that for the Potts model exponentially 
decaying finite-size effects are also possible \cite{citeulike:3612691, 
citeulike:3720308, borgs.kotecky:1992}. The obvious advantage of exponential 
decay is that $\epsilon_\infty$ is approached much faster with increasing $L$, 
compared to the power law decay of \eq{eq:lcv}. This means that moderate system 
sizes may suffice to locate the transition, thereby saving valuable computer 
time. As liquid crystal phase transitions are in any case expensive to 
simulate, such a gain in efficiency would certainly be highly desirable.

We will show in this paper that it is indeed possible to locate first-order IN 
transitions from finite-size simulation data with shifts that vanish much faster 
than $1/L^d$. This is possible by considering $\epsilon_{L,k}$, the inverse 
temperature at which the \ahum{ratio-of-weights} of the isotropic and the 
nematic phases is equal to a value $k$. This ratio-of-weights is obtained from 
the order parameter distribution $P_{L,\epsilon}(\lambda)$, defined as the 
probability to observe an order parameter $\lambda$, when simulating a system of 
size $L$ at inverse temperature $\epsilon$. In the vicinity of the IN transition 
the distribution becomes bimodal, with one peak corresponding to the isotropic 
phase and the other to the nematic phase. The ratio-of-weights is simply the 
ratio of the peak areas. Provided $k$ is chosen optimally $\epsilon_{L,k}$ 
approaches $\epsilon_\infty$ extremely rapidly as $L$ increases, yielding an 
economic alternative over \eq{eq:lcv}. A prerequisite is that the transition 
must be strong enough first-order for the ratio-of-weights to be meaningfully 
calculated \cite{note2}. For this reason, we do not consider the original LL 
model, as the transition is extremely weak here, but a variation of it.

In this paper we firstly provide the details of the modified LL 
model in \sect{model}, together with a description of the 
simulation method that was used to obtain the order parameter 
distribution. Next, we measure $\epsilon_\infty$ using the 
\ahum{standard} approach of extrapolating $\epsilon_{L,CV}$ via 
\eq{eq:lcv}, as well as using the \ahum{new} approach based on 
$\epsilon_{L,k}$. In particular we demonstrate how to locate the optimal 
value $k_{\rm opt}$, along which finite-size effects are minimal. As 
expected, both approaches are in good agreement, with the essential 
difference that $\epsilon_{L,k}$ converges to $\epsilon_\infty$ 
already for very small systems. This fast convergence property was 
observed at all transitions studied by us, irrespective of space and 
spin dimension. We also consider the finite-size scaling of the latent 
heat density and show that, for IN transitions, $k_{\rm opt}$ becomes 
the \ahum{analogue} of the number of Potts states $q$. Finally we 
present a summary of our findings in \sect{conc}.

\section{model and simulation method}
\label{model}

\subsection{Modified LL model}

In order to study finite-size effects at phase transitions, simulation 
data of high statistical quality are essential. This sets a limit on the 
complexity of the models that can be handled as well as on the system 
size. For our purposes already the simple LL model is too demanding, the 
problem being that the IN transition in this model is extremely weak. 
Generally, in computer simulations, first-order phase transitions are 
identified by measuring the probability distribution of the order 
parameter \cite{citeulike:3717210}. At the transition, this distribution 
displays two peaks: one corresponding to the isotropic phase and the 
other to the nematic phase. In the thermodynamic limit, the peaks become 
sharper, and ultimately a distribution of two $\delta$-functions is 
obtained. In finite systems, however, the peaks are broad and possibly 
overlapping, especially when the transition is very weak. Such behaviour 
is observed in the LL model: even in simulation boxes of $L=70$ lattice 
spacings the peaks strongly overlap and the logarithm of the peak 
height, measured with respect to the minimum in between, is less than $2 
\, k_B T$ \cite{citeulike:3740162}. Since the peaks overlap one never 
truly sees pure phases, which complicates the analysis. In order to 
yield reasonable results we require in this paper that the peaks in 
$\ple$ be well-separated. More precisely, it must be possible to assign 
a \ahum{cut-off} separating the peaks, on which the final results may 
not sensitively depend. For this reason we do not consider the original 
LL model but rather a generalization of it, where the exponent $p$ of 
\eq{eqll} exceeds the LL value. We expect this will lead to a much 
stronger first-order IN transition \cite{physrevlett.52.1535, 
physrevlett.89.285702, enter.romano.ea:2006}, so distributions will 
display non-overlapping peaks already in moderately sized systems. In 
fact by using a large exponent $p$ in \eq{eqll} strong first-order IN 
transitions may be realized even in purely two-dimensional systems 
\cite{vink:2006*b, vink.wensink:2007}. Hence, the model that we consider 
is just the LL model of \eq{eqll} but with $p>2$. Note the absolute 
value $| \cdot |$ such that the system is invariant under inversion of 
the spin orientation. We thus impose the symmetry of liquid crystals 
although we believe that our results also apply to magnetic systems.

Note that the use of a large exponent $p$ in \eq{eqll} may also yield a 
better description of experiments on confined liquid crystals. The 
latter systems are quasi two-dimensional. If one studies the LL model in 
two dimensions, i.e.~with $p=2$, and three-dimensional spins, a true 
phase transition appears to be absent \cite{citeulike:3687077}. In 
contrast, experiments clearly reveal that transitions do occur. In fact,
these transitions appear to be of the IN type and are quite strong, as 
manifested by pronounced coexistence between isotropic and nematic 
domains \cite{citeulike:2811025}. Such behaviour cannot be reproduced 
easily with the standard LL model, but it can be using the modified 
version considered in this work, with a sufficiently large exponent~$p$.

\subsection{Transition matrix Wang-Landau sampling}

Following earlier work on the LL model \cite{physrevlett.69.2803, 
citeulike:3740162} our simulations are based on the order-parameter 
distribution. We use the energy $E$ of \eq{eqll} as order parameter and 
aim to measure $\plen$ as accurately as possible. Recall that $\plen$ 
is the probability to observe energy $E$, in a system of size $L$, at 
inverse temperature~$\epsilon$. Depending on the case of interest, the 
simulations are performed on square or cubic lattices of linear size 
$L$, using periodic boundary conditions.

In order to obtain $\plen$ we use Wang-Landau (WL) sampling 
\cite{wang.landau:2001, citeulike:278331} additionally optimized by 
recording some elements of the transition matrix (TM) 
\cite{citeulike:202909, citeulike:3577799}. The aim of WL sampling is to 
perform a random walk in energy space, such that all energies are 
visited equally often. To this end, we use single spin dynamics, whereby 
one of the spins is chosen randomly and given a new random orientation. 
The new state is accepted with probability
\begin{equation}\label{eqwl}
 p(E_I \rightarrow E_J) = \min 
 \left[ \frac{ g(E_I) }{ g(E_J) }, 1 \right],
\end{equation}
with $E_I$ and $E_J$ the energies of the initial and final states 
respectively and $g(E)$ the density of states. The density of states is 
unknown beforehand and $g(E)$ is initially set so $g(E)=1$. Upon 
visiting any particular energy the corresponding density of states is 
multiplied by a modification factor $f \geq 1$. We also keep track of 
the histogram $H(E)$, counting the number of times each energy $E$ is 
visited. Once $H(E)$ contains sufficient information over the range of 
energy of interest, the modification factor $f$ is reduced and the 
energy histogram $H(E)$ is reset to zero. These steps are repeated until 
$f$ has become close to unity, after which changes in the density of 
states become negligible. The sought order parameter distribution is 
then obtained from $\plen \propto g(E) \exp(-\epsilon E)$.

The above procedure is the standard WL algorithm, which works extremely well in 
many cases \cite{physreva.6.426}. However, it has been noted 
\cite{citeulike:3577799, citeulike:3577810} that the WL algorithm in its 
standard form reaches a limiting accuracy, after which the statistical quality 
of the data no longer improves, no matter how much additional computer time is 
invested. Hence, these authors also propose to measure the TM elements $T(E_I 
\to E_J)$. These are defined as the number of times that, being in a
state with energy $E_I$, a state with energy $E_J$ is proposed,
irrespective of whether the new state is accepted. From the TM
elements one can estimate
\begin{equation}
 \Omega(E_I \to E_J) = \frac{ T(E_I \to E_J) }{\sum_K T(E_I \to E_K) },
\end{equation}
which is the probability that being in state with energy $E_I$, a 
move to a state with energy $E_J$ is proposed. This is related to the 
density of states via \cite{citeulike:202909}
\begin{equation}\label{gtm}
 \frac{ g(E_I) }{ g(E_J) } = 
 \frac{ \Omega(E_J \to E_I) }{ \Omega(E_I \to E_J) }.
\end{equation}
Hence, by recording TM elements the density of states 
can also be constructed, the great advantage being that rejected moves 
also give useful information.

To combine WL sampling with the TM method we somewhat follow 
\cite{citeulike:3577799}. At the start of the simulation the density of 
states $g(E)$ is set to unity, while the energy histogram $H(E)$ and the 
TM elements are set to zero. We perform one WL iteration, i.e.~accepting 
moves conform \eq{eqwl}, using a high modification factor $\ln f = 1$. 
At each move both $H(E)$ and the TM elements are updated. We continue to 
simulate until all bins in $H(E)$ contain at least $n$~entries over the 
chosen energy range. We then use the TM elements to construct a new 
density of states, which serves as the starting density of states for 
the next WL iteration. For the next iteration $H(E)$ is reset to zero, 
the modification factor is reduced to $(\ln f) / l$ but the TM elements 
remain untouched. These steps are repeated until $\ln f \approx {\cal 
O}(10^{-20})$, after which we store the corresponding density of states 
$g_P(E)$. This marks the end of the \ahum{prepare} stage.

Next we proceed with the \ahum{collect} stage. The TM~elements are set 
to zero, whereas $H(E)$ is no longer needed. During collection we sample 
according to \eq{eqwl} using $g_P(E)$ as estimate for the density of 
states. However, only the TM~elements and not $g_P(E)$ are further 
updated. As collection proceeds the accuracy of the TM~elements 
increases indefinitely, as does the accuracy of the density of states 
obtained from them. The reason to have a separate \ahum{collect} stage 
is because during \ahum{prepare} detailed balance is not strictly 
obeyed, due to the initially large modification factor $f$ 
\cite{citeulike:3577799}. For this reason, $g_P(E)$ could be biased and 
we are reluctant to perform finite-size scaling with it.

During the \ahum{prepare} stage small values $n \approx 10$ together 
with large values $l \approx 5-10$ can be used. This significantly 
speeds up the simulation and similar observations have been made in 
other works \cite{citeulike:3577799}. Histograms were collected by 
discretizing the energy in bins of resolution $\Delta = 1 \, k_B T$. In 
order to avoid \ahum{boundary effects} during WL sampling states are 
counted as in \cite{citeulike:1247464}. To reduce memory consumption, 
only the nearest-neighbor elements $T(E_I \to E_I \pm \Delta)$ of the TM 
along with the normalization $\sum_K T(E_I \to E_K)$ are stored. Since 
single spin dynamics are used, these are the dominant entries. 
Constructing the density of states using \eq{gtm} and recursion is then 
a straightforward matter. If all TM~elements were to be used, 
constructing the density of states becomes more complex while not 
yielding significantly higher accuracy \cite{citeulike:202909}, so this 
is not attempted here. The required computer time depends sensitively on 
the size of the system. For small systems consisting of $\approx 
1000$~spins, the \ahum{prepare} stage can be completed in as short a 
time as 15~minutes. For larger systems containing $10,000$~spins or more 
this can take more than one week. In these cases it is necessary to 
collect the density of states over a number of separate energy intervals 
with a single processor assigned to each interval. Such a 
parallelization is trivially implemented. The \ahum{collect} phase 
typically lasts as long as the \ahum{prepare} phase except for very 
large systems, where it is found that it takes a much longer time to 
obtain an equivalently accurate density of states.

\section{Results and Analysis}
\label{results}

We have performed extensive simulations of \eq{eqll} varying both the 
space and spin dimension, as well as the exponent $p$. More precisely, 
the following scenarios are considered:
\begin{enumerate}
\item three-dimensional lattices, three-dimensional spins with $p=5-45$,
\item two-dimensional lattices, three-dimensional spins with $p=20-50$ and
\item two-dimensional lattices, two-dimensional spins with $p=150-1000$.
\end{enumerate}
On three-dimensional lattices, it is well accepted that the IN 
transition is first-order. The fact that the IN transition can also be 
first-order in two dimensions is perhaps less well known. In this case, 
first-order transitions only appear provided the exponent $p$ of 
\eq{eqll} is sufficiently large \cite{physrevlett.52.1535, 
physrevlett.89.285702, enter.romano.ea:2006}. Hence, in two dimensions, 
one generally needs $p \gg 2$ in order to observe a first-order 
transition, and it is important to verify that such a transition is 
indeed taking place. If one additionally lowers the spin dimension from 
$3 \to 2$, even greater exponents $p$ are required. For this reason, the chosen
$p$ ranges vary significantly between the three scenarios.

\subsection{Determining the order of the transition}

\begin{figure}
\begin{center}
\includegraphics[width=\figwidth]{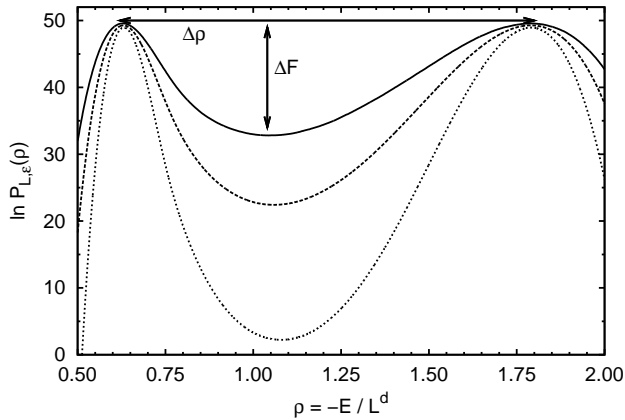}
\caption{\label{fig:pn} Logarithm of $\plen$ using $p=10$ in \eq{eqll} for 
system sizes $L=10,12,15$ (from top to bottom), cubic lattices, and 
three-dimensional spins. In each of the distributions $\epsilon$ was tuned so 
the peaks are of equal height. The barrier $\Delta F$, here marked for the 
$L=10$ system, is defined as the height of the peaks measured with respect to 
the minimum in between. The peak-to-peak distance $\Delta \rho$ corresponds to the 
latent heat density. Note that we have plotted the distributions as a function 
of the negative energy density: the left peak thus corresponds to the isotropic 
phase and the right peak to the nematic phase.}
\end{center}
\end{figure}

\begin{figure}
\begin{center}
\includegraphics[width=\figwidth]{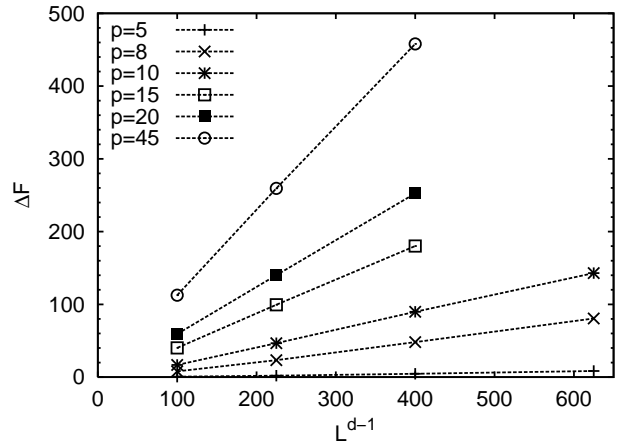}
\caption{\label{fig:dfcube} Variation of $\Delta F$ versus $L^{d-1}$ for 
$d=3$ dimensional lattices, three-dimensional spins, and the various 
values of $p$ as indicated.}
\end{center}
\end{figure}

\begin{figure}
\begin{center}
\includegraphics[width=\figwidth]{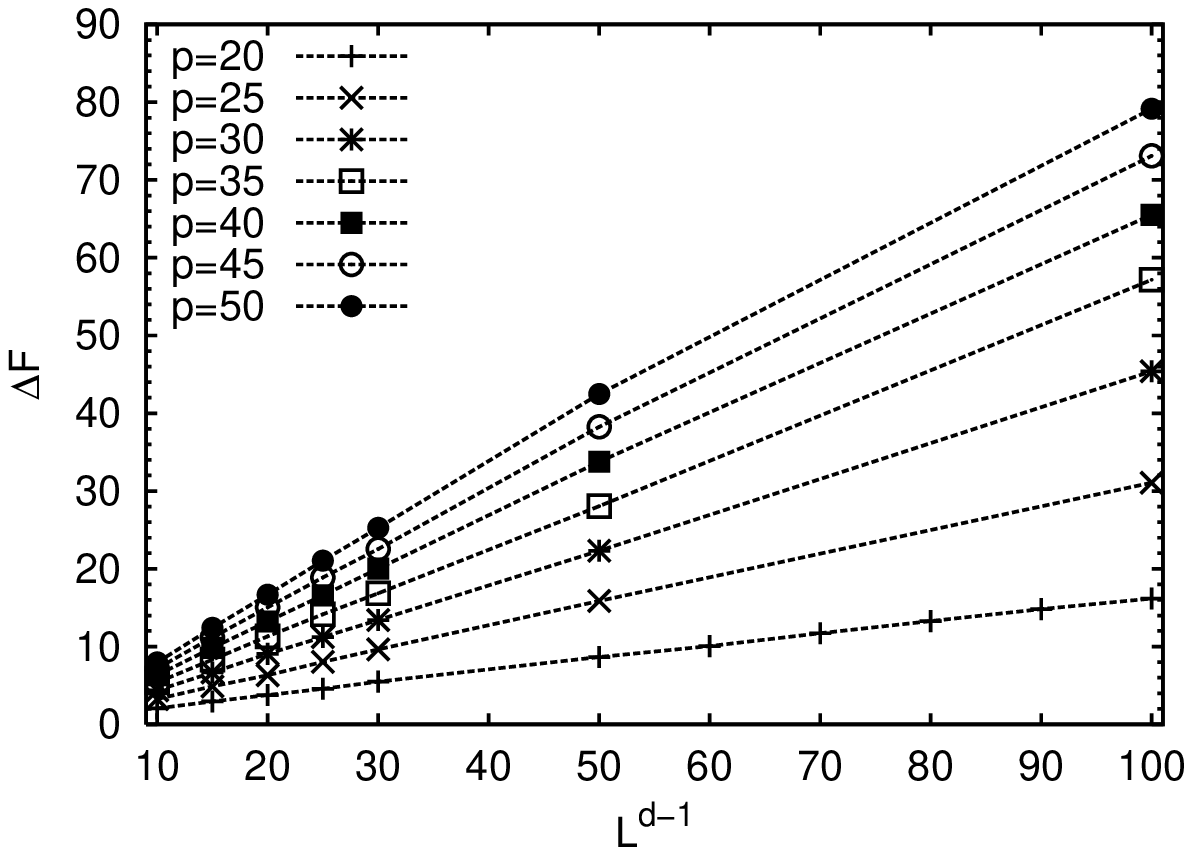}
\caption{\label{fig:df_flat} Similar to \fig{fig:dfcube} but for 
two-dimensional lattices and three-dimensional spins.}
\end{center}
\end{figure}

\begin{figure}
\begin{center}
\includegraphics[width=\figwidth]{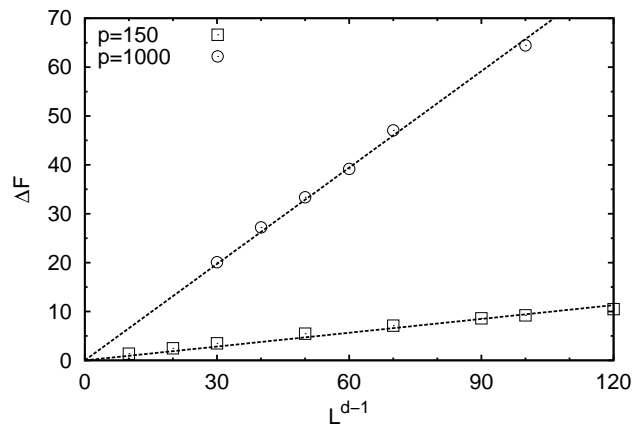}
\caption{\label{fig:df2d} Similar to \fig{fig:dfcube} but for 
two-dimensional lattices and two-dimensional spins.}
\end{center}
\end{figure}

In order to verify the presence of a first-order transition we use the 
scaling method of Lee and Kosterlitz \cite{physrevlett.65.137, 
citeulike:3908342}. Recall that the order parameter distribution becomes 
bimodal in the vicinity of the IN transition; see \fig{fig:pn} for an 
example. The idea of Lee and Kosterlitz is to monitor the peak heights 
$\Delta F$ of the logarithm of the order parameter distribution, measured 
with respect to the minimum \ahum{in-between} the peaks. At a first-order 
transition $\Delta F$ corresponds to the formation of interfaces between 
coexisting isotropic and nematic domains \cite{binder:1982}. In $d$ 
spatial dimensions it is therefore expected that $\Delta F \propto 
L^{d-1}$, providing a straightforward recipe to identify the
transition type: a linear increase of $\Delta F$ versus $L^{d-1}$
indicates that a first-order transition is taking place, with the 
slope yielding the interfacial tension \cite{binder:1982}, whereas for 
a continuous transition $\Delta F$ becomes independent of $L$, or 
vanishes altogether if no transition takes place at all in the 
thermodynamic limit. In \fig{fig:dfcube} $\Delta F$ is plotted for
the purely three-dimensional case; the linear increase is clearly 
visible, confirming the presence of a first-order transition. 
For two-dimensional lattices the results have been collected in 
\fig{fig:df_flat} and \fig{fig:df2d}. Once again the presence of a 
first-order transition is confirmed. Note that on two-dimensional 
lattices the slopes of the lines correspond to line tensions.

\subsection{Extrapolation of $\epsilon_{L,CV}$}

\begin{figure}
\begin{center}
\includegraphics[width=\figwidth]{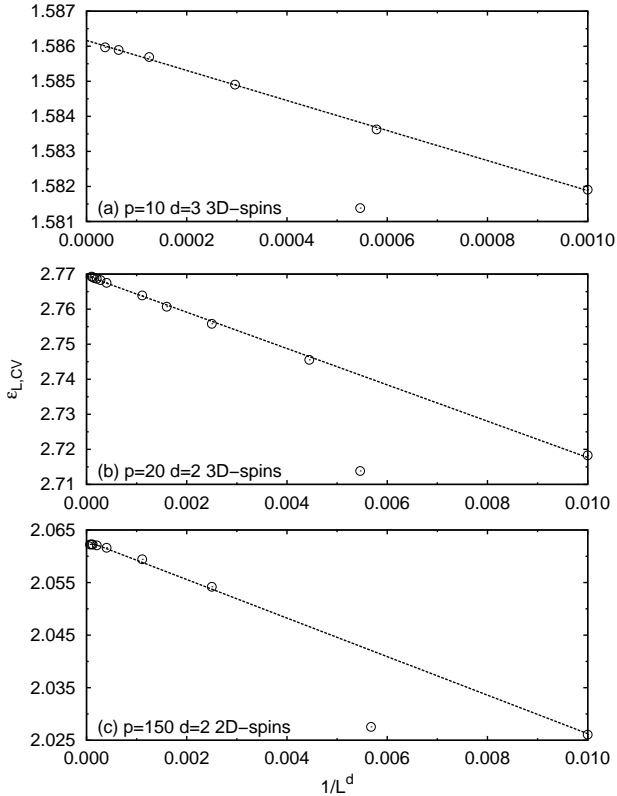}
\caption{\label{fig:cv} Estimation of $\epsilon_\infty$ via 
extrapolation of $\epsilon_{L,CV}$ using \eq{eq:lcv}. Shown is 
$\epsilon_{L,CV}$ versus $1/L^d$ using (a) $p=10$, cubic lattices and 
three-dimensional spins, (b) $p=20$, square lattices and 
three-dimensional spins and (c) $p=150$, square lattices and 
two-dimensional spins. The open symbols are simulation data and 
the lines are fits to \eq{eq:lcv}.}
\end{center}
\end{figure}

\begin{table*}

\caption{\label{tab3d} Properties of the IN transition of \eq{eqll} for 
three-dimensional lattices and three-dimensional spins versus $p$. Listed are 
the fit parameters $\epsilon_{\infty,CV}$ and $\alpha$ of \eq{eq:lcv}, the best 
estimate $\epsilon_{\infty,k}$ obtained from the convergence of $\epsilon_{L,k}$ 
along $k_{\rm opt}$, the logarithm of $k_{\rm opt}$ with uncertainty $\Delta k$, 
the latent heat density ${\cal L}_\infty$ and the ratio $\ln k_{\rm opt} / 
{\cal L}_\infty$.}

\begin{ruledtabular}
\begin{tabular}{c|cc|cc|cc}
$p$ & $\epsilon_{\infty,CV}$ & $\alpha$ & $\epsilon_{\infty,k}$ & $\ln k_{\rm opt} \pm \Delta k$ & $\cal L_\infty$ & $\ln k_{\rm opt} / {\cal L}_\infty$ \\ \hline
  5 & 1.3969 & 6.62 & $1.3970 \pm 0.001$ & $2.7 \pm 0.6$ & $0.358 \pm 0.010$ & $5.9-9.2$ \\
  8 & 1.5207 & 5.28 & $1.5207 \pm 0.001$ & $5.0 \pm 0.2$  & $0.908 \pm 0.005$ & $5.3-5.7$ \\
 10 & 1.5862 & 4.28 & $1.5864 \pm 0.001$ & $5.3 \pm 0.5$ & $1.156 \pm 0.005$ & $4.2-5.0$ \\
 15 & 1.7126 & 3.94 & $1.7126 \pm 0.001$ & $6.0 \pm 0.2$ & $1.523 \pm 0.002$ & $3.8-4.1$ \\
 20 & 1.8063 & 3.76 & $1.8063 \pm 0.001$ & $6.4 \pm 0.3$ & $1.727 \pm 0.002$ & $3.5-3.9$ \\
 45 & 2.0838 & 3.55 & $2.0838 \pm 0.001$ & $8.1 \pm 1.0$ & $2.120 \pm 0.001$ & $3.3-4.3$ \\
\end{tabular}
\end{ruledtabular}
\end{table*}

\begin{table*}
\caption{\label{tabq2d} Similar to \tab{tab3d} but for two-dimensional 
lattices and three-dimensional spins.}
\begin{ruledtabular}
\begin{tabular}{c|cc|cc|cc}
$p$ & $\epsilon_{\infty,CV}$ & $\alpha$ & $\epsilon_{\infty,k}$ & $\ln k_{\rm opt} \pm \Delta k$ & $\cal L_\infty$ & $\ln k_{\rm opt} / {\cal L}_\infty$ \\ \hline
 20 & 2.7695 & 5.18 & $2.7698 \pm 0.001$ & $4.1 \pm 0.4$ & $0.7145 \pm 0.001$ & $5.2-6.3$ \\
 25 & 2.8678 & 4.84 & $2.8679 \pm 0.001$ & $4.5 \pm 0.2$ & $0.8900 \pm 0.001$ & $4.8-5.3$ \\
 30 & 2.9517 & 4.69 & $2.9517 \pm 0.001$ & $4.8 \pm 0.2$ & $1.0023 \pm 0.001$ & $4.6-5.0$ \\
 35 & 3.0240 & 4.58 & $3.0241 \pm 0.001$ & $5.2 \pm 0.6$ & $1.0832 \pm 0.001$ & $4.2-5.4$ \\
 40 & 3.0882 & 4.52 & $3.0882 \pm 0.001$ & $5.2 \pm 0.2$ & $1.1432 \pm 0.001$ & $4.4-4.7$ \\
 45 & 3.1456 & 4.47 & $3.1455 \pm 0.001$ & $5.2 \pm 0.5$ & $1.1936 \pm 0.001$ & $3.9-4.8$ \\
 50 & 3.1976 & 4.50 & $3.1976 \pm 0.001$ & $5.6 \pm 0.5$ & $1.2320 \pm 0.001$ & $4.1-5.0$ \\
\end{tabular}
\end{ruledtabular}
\end{table*}

\begin{table*}
\caption{\label{tab2d} Similar to \tab{tab3d} but for two-dimensional 
lattices and two-dimensional spins.}
\begin{ruledtabular}
\begin{tabular}{c|cc|cc|cc}
$p$ & $\epsilon_{\infty,CV}$ & $\alpha$ & $\epsilon_{\infty,k}$ & $\ln k_{\rm opt} \pm \Delta k$ & $\cal L_\infty$ & $\ln k_{\rm opt} / {\cal L}_\infty$ \\ \hline
150  & 2.063 & 3.67 & $2.0628 \pm 0.0005$ & $1.9 \pm 0.8$ & $0.648 \pm 0.005$ & $1.7-4.2$ \\
1000 & 2.486 & 3.31 & $2.4865 \pm 0.0006$ & $4.8 \pm 1.3$ & $1.270+0.005$ & $2.8-4.8$ \\
\end{tabular}
\end{ruledtabular}
\end{table*}

Next, we measure the thermodynamic limit inverse temperature 
$\epsilon_\infty$ by means of extrapolating $\epsilon_{L,CV}$ via 
\eq{eq:lcv}. Recall that $\epsilon_{L,CV}$ is the finite-size inverse 
temperature where the specific heat
\begin{equation}\label{eqsph}
 C_L = \frac{ \avg{E^2} - \avg{E}^2 }{ L^d }
\end{equation}
attains its maximum. Shown in \fig{fig:cv}(a) is $\epsilon_{L,CV}$ versus 
$1/L^d$ for the purely three-dimensional case using $p=10$ - results for 
different $p$ are qualitatively similar and therefore not explicitly shown. In 
agreement with earlier simulations of the \ahum{original} LL model 
\cite{physrevlett.69.2803, citeulike:3740162}, the data are well described by 
\eq{eq:lcv} and from the fit $\epsilon_\infty$ can be meaningfully obtained. 
The resulting fit parameters are collected in \tab{tab3d}. Repeating the same 
analysis for two-dimensional lattices yields similar results; some typical plots 
are shown in \fig{fig:cv}(b) and~(c) with the resulting fit parameters 
collected in Tables~II and~III.

\subsection{Extrapolation of $\epsilon_{L,k}$}

We now arrive at the main result of this paper, namely the estimation of 
$\epsilon_\infty$ by monitoring $\epsilon_{L,k}$. Recall that 
$\epsilon_{L,k}$ is defined as the finite-size inverse temperature where 
the equality
\begin{equation}\label{mm}
  W_N / W_I = k
\end{equation}
is obeyed, with $W_N$ and $W_I$ the areas under the nematic and 
isotropic peaks of the order parameter distribution respectively. No 
matter what value of $k$ is used, provided it is positive and finite, we 
expect that $\lim_{L \to \infty} \epsilon_{L,k} = \epsilon_\infty$. The 
reason is that in the thermodynamic limit a bimodal order parameter 
distribution survives only at $\epsilon_\infty$ and not anywhere else 
\cite{citeulike:3908342}. Hence, keeping the area ratio fixed at some 
value of $k$ whilst increasing $L$, $\epsilon_{L,k}$ will definitely 
approach $\epsilon_\infty$. The rate of the convergence, however, does 
depend on $k$. Assuming that the prediction of Borgs and Kotecky for the 
Potts model also holds at first-order IN transitions, it should be 
possible to locate an optimal value $k_{\rm opt}$ at which the 
convergence to $\epsilon_\infty$ is fastest and hopefully faster than 
$1/L^d$. Therefore, we propose to manually inspect the convergence of 
$\epsilon_{L,k}$ using several values of $k$.

A prerequisite for numerically solving \eq{mm} is that the transition 
must be sufficiently first-order in order for bimodal distributions 
$\plen$ with well-separated peaks to appear. By this we mean that the 
barrier $\Delta F$, defined in \fig{fig:pn}, is large enough. The areas 
of the nematic and isotropic peaks may then be calculated using
\begin{equation}
 W_N = \int_{-\infty}^{E_c} \plen \, dE, \hspace{2mm}
 W_I = \int_{E_c}^0 \plen \, dE,
\end{equation}
where we remind the reader that the energy in our model is negative. The 
details of defining the \ahum{cut-off} energy $E_c$ are somewhat 
arbitrary, but as states around $E_c$ contribute exponentially little 
to the peak areas, the precise form does not matter 
\cite{citeulike:3596258}. In this work $E_c$ is taken to be the average 
$E_c = \int E \plen \, dE$, with $\plen$ obtained at equal-height, 
i.e.~as in \fig{fig:pn}. Once $E_c$ has been set its value is kept fixed 
whilst solving \eq{mm}.

\begin{figure}
\begin{center}
\includegraphics[width=\figwidth]{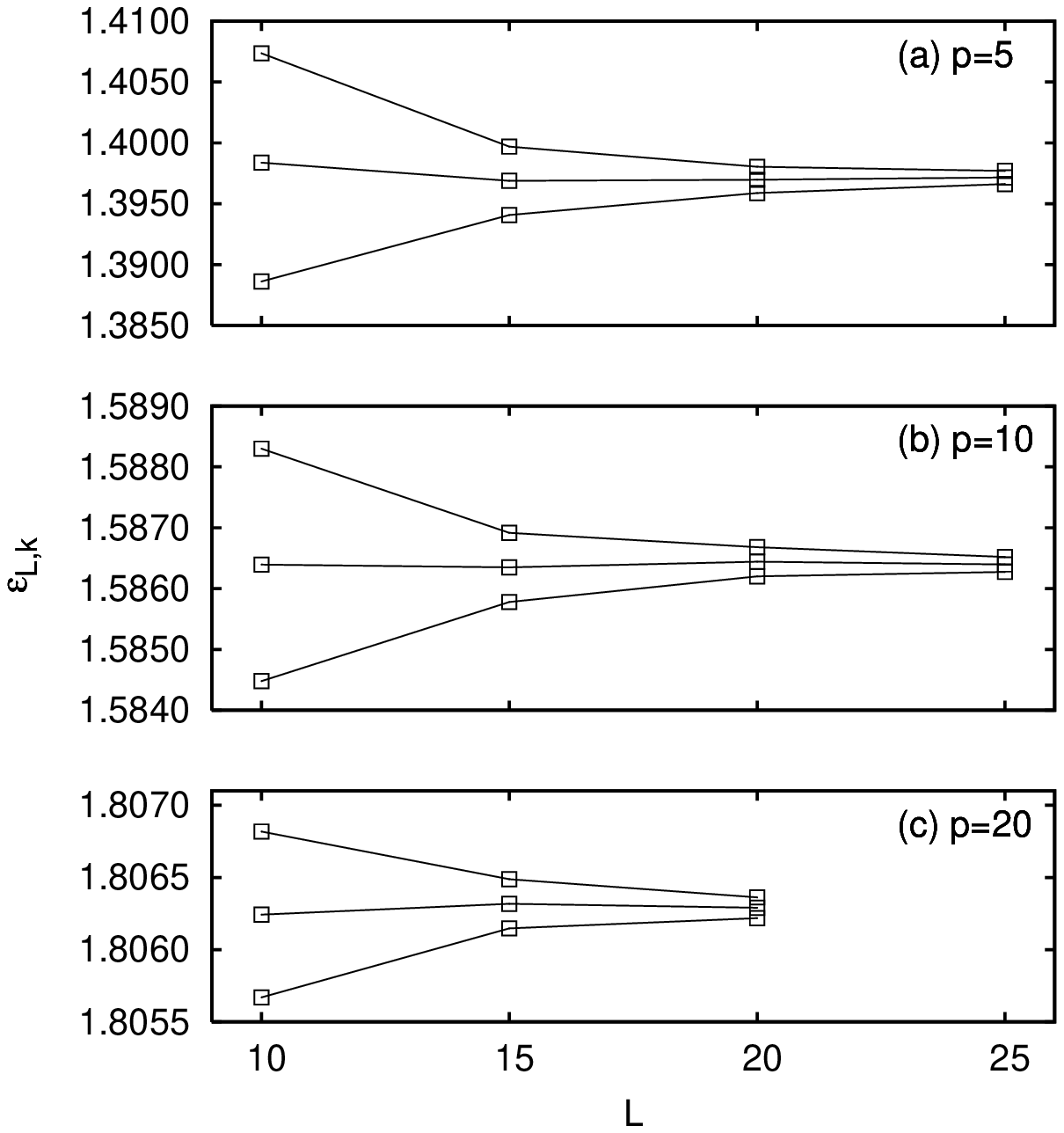}

\caption{\label{fig:multicube} Variation of $\epsilon_{L,k}$ versus $L$, for 
three-dimensional lattices and three-dimensional spins, using different 
exponents $p$ as indicated. The symbols are simulation data, the lines serve to 
guide the eye. The central curves in each plot show $\epsilon_{L,k}$ using $k = 
k_{\rm opt}$ along which finite-size effects are minimal; also shown is 
$\epsilon_{L,k}$ using $k = k_{\rm opt} - 5 \Delta k$ (lower curves), and $k = 
k_{\rm opt} + 5 \Delta k$ (upper curves). The methods for locating $k_{\rm opt}$ 
and $\Delta k$ are explained in the text and the resulting values, as
well as the estimates of $\epsilon_\infty$, are listed in \tab{tab3d}.}

\end{center}
\end{figure}

\begin{figure}
\begin{center}
\includegraphics[width=\figwidth]{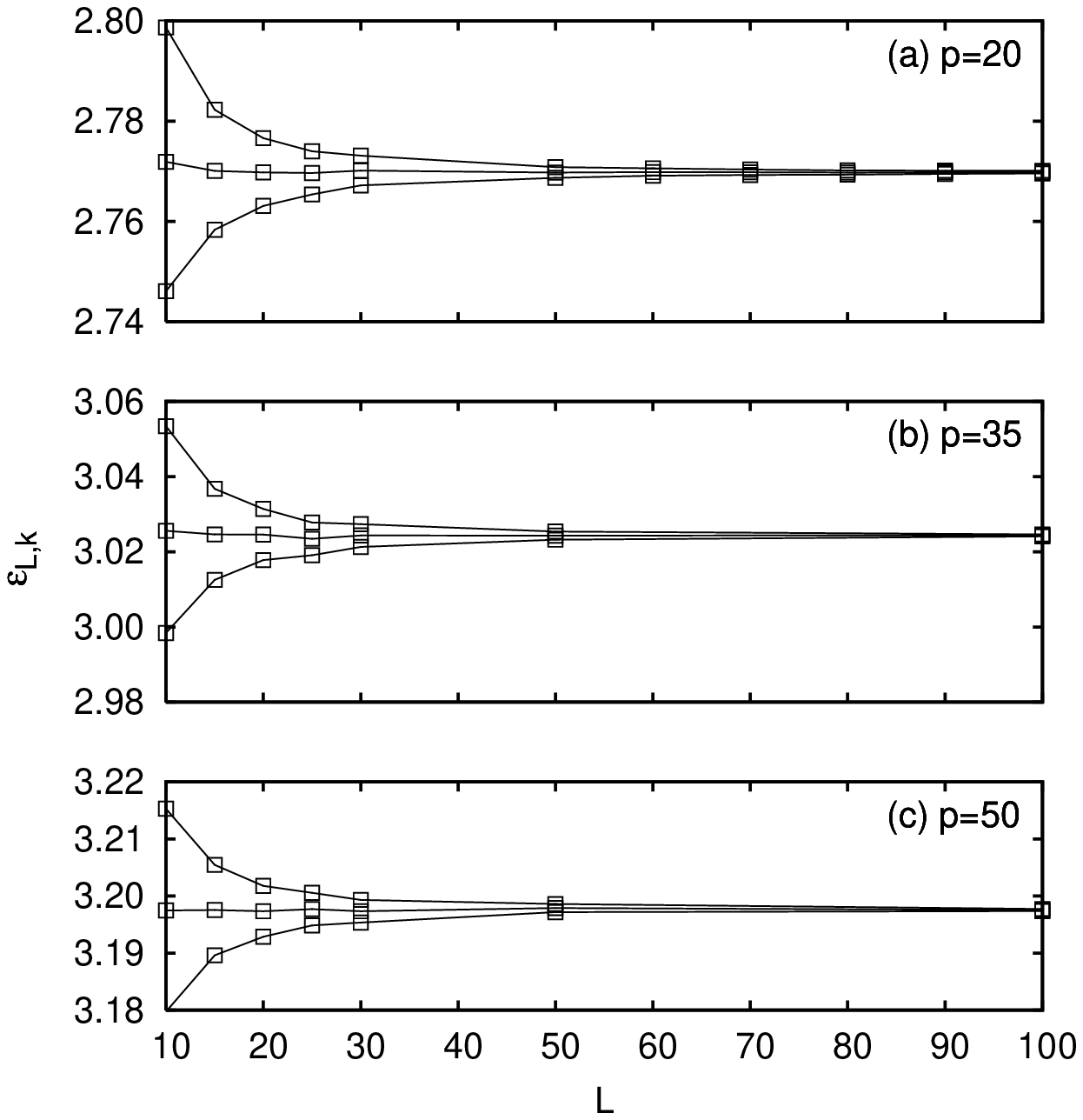} 
\caption{\label{fig:multi_flat} Similar to \fig{fig:multicube} but for 
two-dimensional lattices and three-dimensional spins. Numerical 
estimates are given in \tab{tabq2d}.}
\end{center}
\end{figure}

\begin{figure}
\begin{center}
\includegraphics[width=\figwidth]{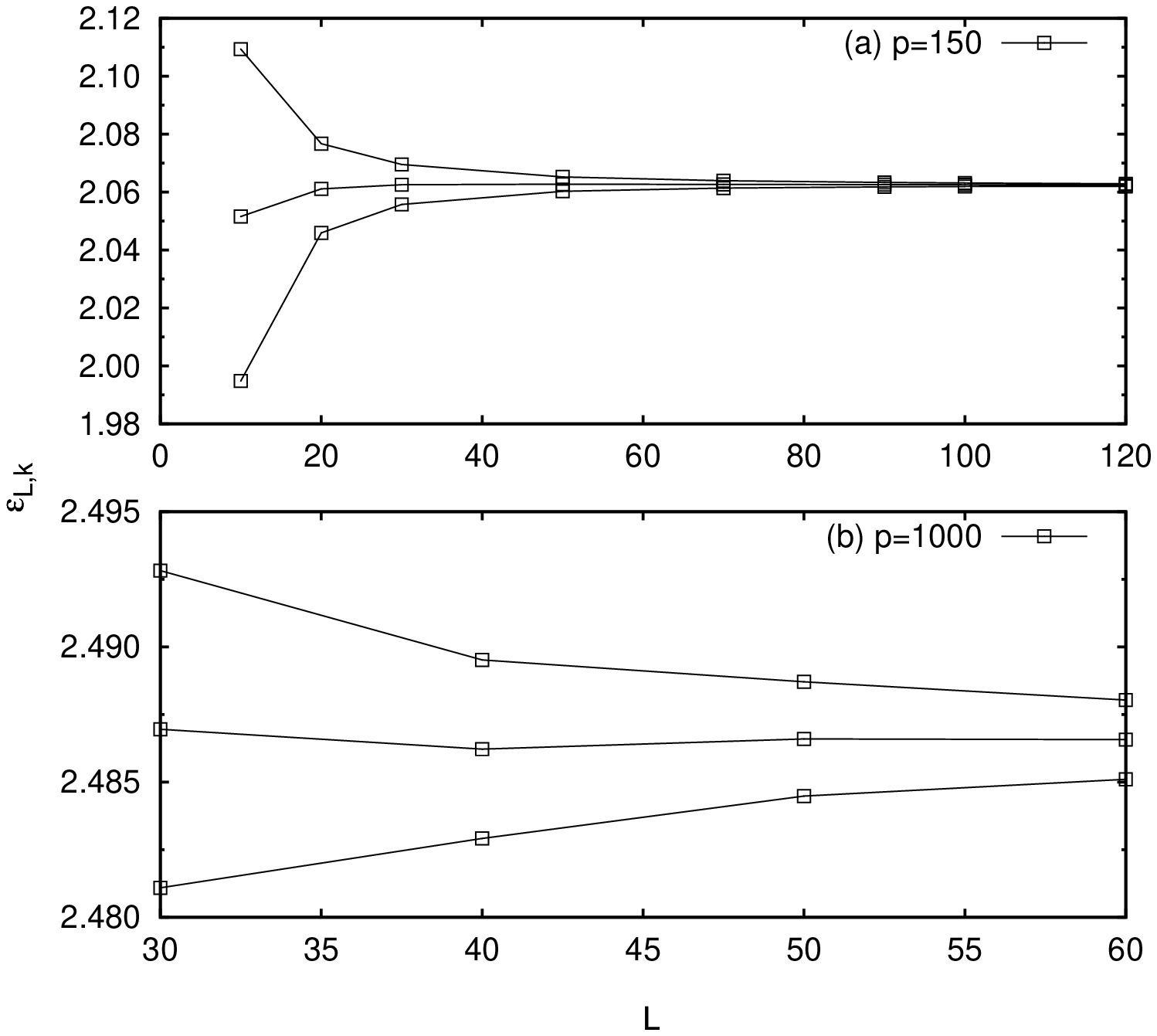}
\caption{\label{fig:k2d} Similar to \fig{fig:multicube}, but for 
two-dimensional lattices and two-dimensional spins. Numerical estimates 
are given in \tab{tab2d}.}
\end{center}
\end{figure}

For the purely three-dimensional case, the behaviour of $\epsilon_{L,k}$ is shown 
in \fig{fig:multicube}. Using a number of exponents $p$ in \eq{eqll}, we have 
plotted $\epsilon_{L,k}$ versus $L$ for several values of $k$. The data are 
consistent with the expectation that, regardless of $k$, $\epsilon_{L,k}$ 
converges to a common value, corresponding to $\epsilon_\infty$. Note also that 
$\epsilon_\infty$ is approached from above for large $k$, and from below for low 
$k$. Hence, we can indeed identify an optimal value $k_{\rm opt}$ along which 
finite-size effects are minimal. The optimum can be estimated by locating, for a 
pair of system sizes $L_i$ and $L_j$, the inverse temperature $\epsilon_{ij}$ 
where for both system sizes the same ratio $k_{ij}$ of the peak areas is 
observed. By considering all available pairs of system sizes, the average and 
root-mean-square fluctuation in $\epsilon_{ij}$ and $k_{ij}$ can be calculated, 
which then yield $\epsilon_\infty$ and $k_{\rm opt}$ with uncertainties, shown in 
\tab{tab3d}. Although $k_{\rm opt}$ itself is not known very precisely, since 
$\Delta k$ is quite large, very accurate estimates of $\epsilon_\infty$ can still be 
obtained as this quantity is rather insensitive to the precise value of $k_{\rm 
opt}$ being used. This means that the series $\epsilon_{L,k}$ also provides a 
valid method for locating IN transitions. The corresponding estimates of 
$\epsilon_\infty$ are in good agreement with those obtained via extrapolation 
of $\epsilon_{L,CV}$, as inspection of the various tables indicates. The 
practical advantage of using $\epsilon_{L,k}$ with $k = k_{\rm opt}$ is that 
the $L$-dependence is very weak, so much so that $\epsilon_\infty$ 
is captured already in small systems. Similar findings were obtained 
using two-dimensional lattices, of which some typical plots are provided in 
\fig{fig:multi_flat} and \fig{fig:k2d} with the corresponding
numerical estimates collected in Tables~II and~III.

\begin{figure}
\begin{center}
\includegraphics[width=\figwidth]{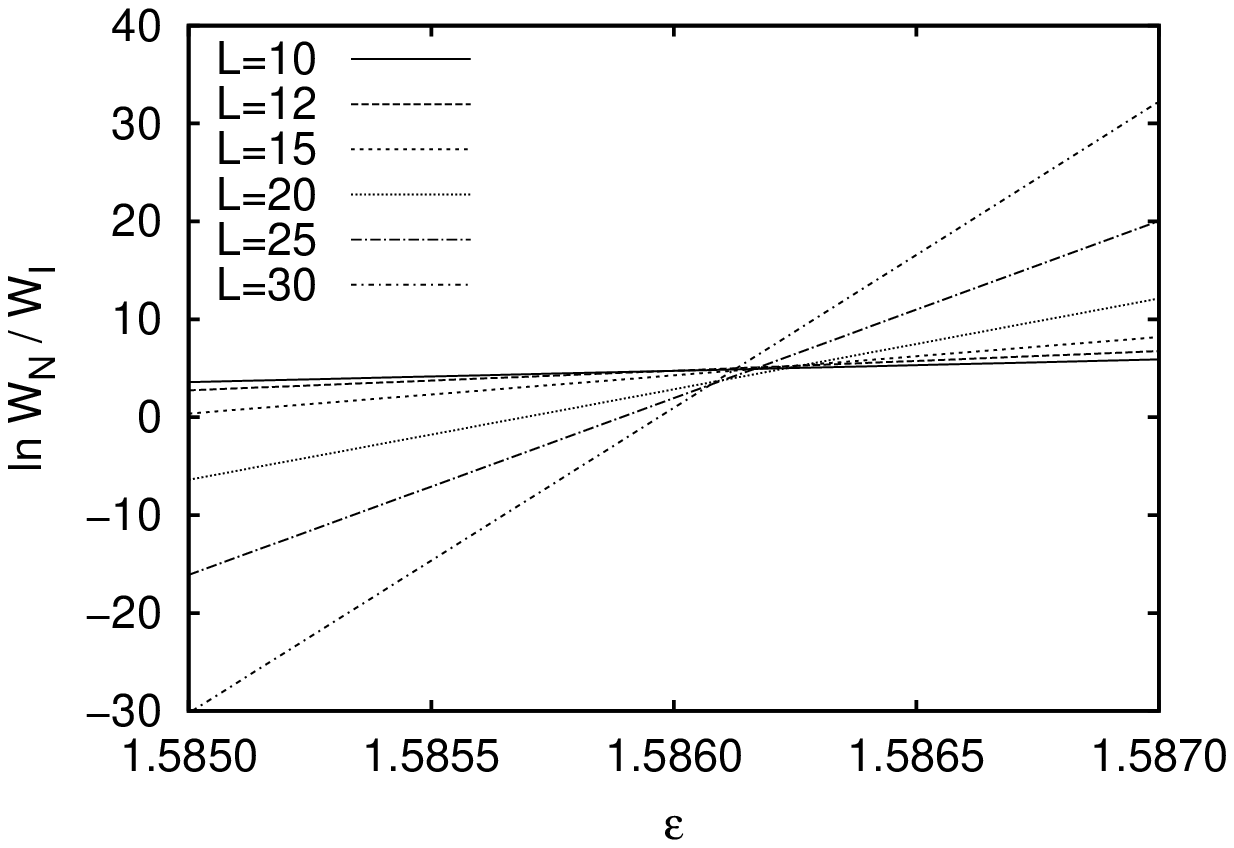}
\caption{\label{fig:jj} Plot of $\ln W_N / W_I$ versus $\epsilon$ for several 
system sizes $L$. This plot was generated using $p=10$ in \eq{eqll}, 
three-dimensional lattices and three-dimensional spins.}
\end{center}
\end{figure}

For non-optimal values $k \neq k_{\rm opt}$, we observe that the shift 
$\epsilon_\infty - \epsilon_{L,k} \propto 1/L^d$, i.e.~the shift vanishes as a 
power law in the inverse volume, similar to $\epsilon_{L,CV}$. At the optimal 
value $k=k_{\rm opt}$, finite-size effects in $\epsilon_{L,k}$ are typically too 
small in order for a meaningful fit to be carried out. Hence, our data confirm 
Borgs and Kotecky in the sense that optimal estimators can be defined which 
converge onto $\epsilon_\infty$ faster than $1/L^d$; whether the optimal 
convergence is indeed exponential requires more accurate data, which is 
currently beyond our reach \cite{note3}.

An alternative, but completely equivalent, method to investigate the convergence 
of $\epsilon_{L,k}$ is presented in \cite{citeulike:3596176, citeulike:3596258, 
citeulike:3608409, citeulike:3610966}, albeit for the Potts model. The idea is to plot 
the area ratio $W_N / W_I$ versus $\epsilon$ for several system sizes. The 
resulting curves are expected to reveal an intersection point at the transition 
inverse temperature; the value of the area ratio at the intersection then 
yields $k_{\rm opt}$. For completeness we have prepared one such plot, see 
\fig{fig:jj}. The curves indeed intersect and give estimates of 
$\epsilon_\infty$ and $k_{\rm opt}$ that are fully consistent with those 
reported in Table~I.

\subsection{Latent heat density}

It appears that scaling relations derived for the Potts model also work 
remarkably well at IN transitions. In agreement with earlier simulations 
of the LL model \cite{physrevlett.69.2803, citeulike:3740162}, the 
validity of \eq{eq:lcv} is confirmed additionally by us. Furthermore, 
our data suggest that an analogue of the Borgs and Kotecky prediction, 
namely that finite-size effects vanish faster than $1/L^d$ at 
appropriate points, can be defined. In this case, it is needed to 
measure $\epsilon_{L,k}$ using the optimal value $k=k_{\rm opt}$. In the 
Potts model it holds that $k_{\rm opt} = q$, where $q$ is the number of 
Potts states. In other words, finite-size effects in the Potts model are 
minimized when the ratio of the peak areas in the order parameter 
distribution is held fixed at $q$. Based on our results, it seems 
reasonable to assume that scaling relations for the Potts model also 
hold at IN transitions, but with $q$ replaced by $k_{\rm opt}$.

\begin{figure}
\begin{center}
\includegraphics[width=\figwidth]{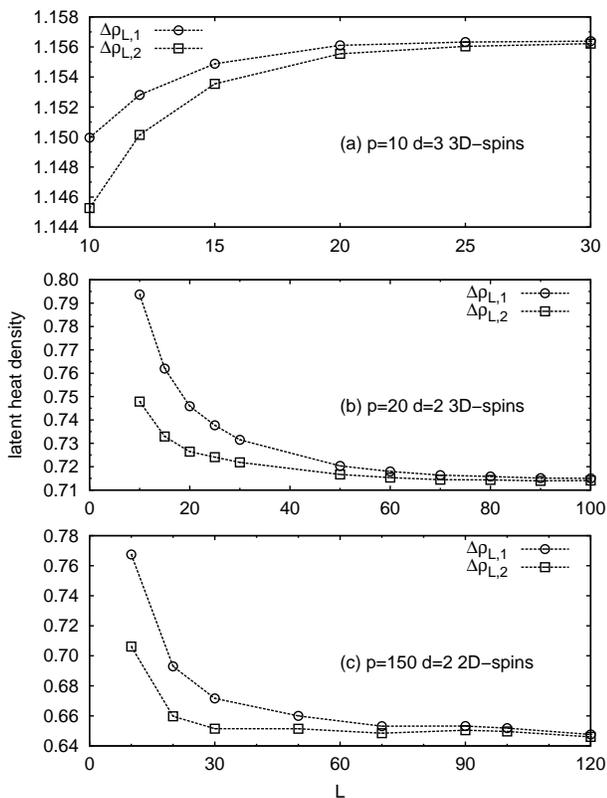}
\caption{\label{fig:lh} Finite-size variation of the latent heat density estimators 
$\Delta \rho_{L,1}$ and $\Delta \rho_{L,2}$. Results are shown for \eq{eqll} using (a) 
$p=10$, three-dimensional lattices and three-dimensional spins, (b) $p=20$, 
two-dimensional lattices and three-dimensional spins and (c) $p=150$, 
two-dimensional lattices and two-dimensional spins.}
\end{center}
\end{figure}

To test this assumption we consider the proportionality constant $\alpha$ from 
\eq{eq:lcv}, which is given by \eq{eq:alpha} for the Potts model. If $q$ can be 
replaced by $k_{\rm opt}$, $\alpha$ should correspond to $\ln k_{\rm opt} / 
{\cal L}_\infty$, where ${\cal L}_\infty$ is the latent heat density. The latter can 
be obtained independently from $C_{L,\rm max} = {\cal L}_\infty^2 L^d / 4$, where 
$C_{L,\rm max}$ is the maximum value of the specific heat in a finite system of 
size $L$ \cite{citeulike:3720308}. Hence, we introduce the latent heat estimator
\begin{equation}
 \Delta \rho_{L,1} = \sqrt{ 4 C_{L, \rm max} / L^d },
\end{equation}
which should approach ${\cal L}_\infty$ as $L \to \infty$. Additionally, 
the latent heat density can be read-off directly, as the peak-to-peak 
distance in the energy distribution, marked $\Delta \rho$ in \fig{fig:pn}. 
Numerically this is expressed by ${\cal M}_L = 2 \avg{|E - \avg{E}|} / 
L^d$; plotting ${\cal M}_L$ versus $\epsilon$ gives a maximum $\Delta 
\rho_{L,2}$, which in the limit $L \to \infty$ also approaches ${\cal 
L}_\infty$. Typical behaviour of $\Delta \rho_{L,i}$ is shown in 
\fig{fig:lh}. As expected, both latent heat estimators converge to a 
common value, which can be read-off reasonably accurately; the resulting 
estimates of ${\cal L}_\infty$ are given in the various tables. Note 
also that ${\cal L}_\infty$ is approached from below in three 
dimensions, whereas in two dimensions, it is approached from above. If 
an appropriate number of two-dimensional lattice layers stacked on top 
of each other were simulated, it is likely that a cross-over regime 
could be found where $\Delta \rho_{L,i}$ depends only weakly on $L$, as 
these systems are effectively in-between two and three dimensions.

Having measured ${\cal L}_\infty$, the ratio $\ln k_{\rm opt} / {\cal 
L}_\infty$ is easily obtained, which may then be compared to $\alpha$, 
see Tables~I-III. The uncertainty is admittedly rather large, but within 
numerical precision, and the relation $\ln k_{\rm opt} / {\cal L}_\infty 
\sim \alpha$ appears to hold.

\section{Summary}
\label{conc}

In this paper we have presented simulation data of first-order 
isotropic-to-nematic transitions in lattice liquid crystals with 
continuous orientational degrees of freedom for various space and 
spin dimensions. As with earlier simulations of this type 
\cite{physrevlett.69.2803, citeulike:3740162}, we find that the 
extrapolation of the finite-size inverse temperature of the specific 
heat maximum $\epsilon_{L,CV}$ can be consistently performed assuming 
a leading $\alpha/L^d$ dependence, exactly as in the Potts model. 
Inspired by this result, we have investigated an alternative approach 
to locate the transition inverse temperature using estimators 
$\epsilon_{L,k}$, defined as the finite-size inverse temperature where 
the ratio of peak areas in the energy distribution is equal to $k$. 
In agreement with the Potts model, $\epsilon_{L,k}$ converges to 
$\epsilon_\infty$ much faster than $1/L^d$, provided an optimal value 
$k=k_{\rm opt}$ is used. Moreover, the ratio $k_{\rm opt} / {\cal  L}_\infty$, 
with ${\cal L}_\infty$ the latent heat density, is remarkably consistent 
with the proportionality constant $\alpha$ from the scaling of 
$\epsilon_{L,CV}$. This leads us to conclude that finite-size 
scaling predictions originally proposed for first-order transitions 
in the Potts model remain valid at first-order IN transitions too, 
but with the number of Potts states $q$ replaced by $k_{\rm opt}$.

It is perhaps somewhat surprising that a continuous spin model at a 
first-order transition, such as the LL model, scales in the same way as 
the Potts model, which is, after all, a discrete spin model. In fact, 
Borgs and Kotecky have remarked that the derivation of their scaling 
results cannot be easily extended to continuous spin models 
\cite{borgs.kotecky:1992}. Nevertheless, the LL model and its variants 
may be more closely connected to the Potts model than one may initially 
think. Note that for large $p$ the Hamiltonian of \eq{eqll} becomes 
increasingly Potts-like, in the sense that the pair interaction 
approaches a $\delta$-function: $\lim_{p \to \infty} | \vec{d}_i \cdot 
\vec{d}_j |^p = \delta(\vec{d}_i, \vec{d}_j)$. This implies that 
neighbouring spins only interact when they are closely aligned and are 
otherwise indifferent to each other, just as in the Potts model. It has 
indeed been suggested that such models approximately resemble $q$-state 
Potts models, with $q \sim \sqrt{p}$ \cite{physrevlett.52.1535}. The 
observed trends in this work are certainly consistent with this 
interpretation. For all cases considered the strength of the transition 
increases with $p$, as manifested by the growing latent heats and 
interfacial tensions, exactly as in the Potts model with increasing $q$. 
Also the upward shift of $\epsilon_\infty$ with $p$ is consistent with 
the Potts model. However, it is clear that new theoretical approaches 
are needed to fully understand finite-size effects at first-order 
transitions in the models studied here. We hope that the present 
simulation results may inspire such efforts.

\acknowledgments

This work was supported by the {\it Deutsche Forschungsgemeinschaft} 
under the Emmy Noether program (VI~483/1-1). We thank Marcus M\"uller 
for stimulating discussions.

\bibliography{mc1975,notes}

\begin{thebibliography}{10}
\bibitem{note2}
Strictly speaking, \eq{eq:lcv} also requires a strong enough first-order
  transition. For example, in the derivation of \olcite{citeulike:3716330}, it
  is assumed that the order parameter distribution consists of two
  non-overlapping Gaussians.
\bibitem{note3}
For the Potts model, the exponential size dependence has been accurately
  resolved \cite{citeulike:3610966}.
\end{thebibliography}

\end{document}